\documentclass[a4paper,11pt]{iopart}
\usepackage{graphicx}
\usepackage{cite}
\usepackage{color}
\usepackage{amsfonts}
\usepackage{srcltx}
\usepackage{color}
\usepackage{amsfonts}
\usepackage{soul}
\usepackage{cancel}
\usepackage{ulem}

\begin{document}

\title{On the symmetry of current probability distributions in jump processes}
\author{A C Barato$^{1,2}$ and R Chetrite$^3$}
\address{
$^1$ II. Institut f\"ur Theoretische Physik, Universit\"at Stuttgart\\ 
Stuttgart 70550, Germany\\
$^2$The Abdus Salam International Centre for Theoretical Physics\\
Trieste 34014, Italy\\
$^3$  Laboratoire J. A. Dieudonn\'e, UMR CNRS 6621, Universit\'e de Nice Sophia-Antipolis\\
 Parc Valrose, 06108 Nice Cedex 02, France
    }

\ead{barato@theo2.physik.uni-stuttgart.de}

\def\ex#1{\langle #1 \rangle}

\begin{abstract}
We study the symmetry of large deviation functions associated with time-integrated currents in Markov pure jump processes. One current known to have a
symmetric large deviation function is the fluctuating entropy production and this is the content of the fluctuation theorem. Here we obtain a necessary condition in order to have a current different from entropy with this symmetry. This condition is related to degeneracies in the  set of increments associated with fundamental cycles from Schnakenberg network theory. Moreover, we consider  four-state systems where we explicitly show that non-entropic time-integrated currents can be symmetric. We also show that these new symmetries, as is the case of the fluctuation theorem, are related to time-reversal. However, this becomes apparent only when stochastic trajectories are appropriately grouped together.  
\end{abstract}

\section{Introduction}

Large deviation theory \cite{Varadhan,ellis85,hollander,touchette09} is the branch of mathematics that deals with exponentially decaying probabilities. Therefore, it is the appropriate mathematical theory for statistical physics. For example, two basic concepts in large deviation theory are that of a rate function (or large deviation function), which gives the rate at which a certain probability distribution decays exponentially, and the scaled cumulant generating function, which is the Legendre-Fenchel transform of the rate function. In equilibrium statistical physics, the microcanonical entropy is a rate function and the canonical free energy is the corresponding scaled cumulant generating function. Moreover, it can be shown that the maximum entropy and minimum free energy principles are consequences of the more general contraction principle, which is central in large deviation theory.

Furthermore, large deviation theory also plays a prominent role for systems out of equilibrium. A series of relations known as fluctuation relations \cite{evans93,evans94,gallavotti95,gallavotti952,jarzynski97,kurchan98,lebowitz99,maes99,crooks99,hatano01,Jia1,seifert05,andrieux07,andrieux07',harris07,kurchan07} are the most general statements known in nonequilibrium statistical physics, for which a general theoretical framework still lacks. Among these relations are the Jarzynski relation \cite{jarzynski97}, the Crooks relation \cite{crooks99}, the Hatano-Sasa relation \cite{hatano01} and the Gallavotti-Cohen-Evans-Morriss (GCEM) fluctuation theorem \cite{evans93,evans94,gallavotti95,gallavotti952}. These statements are about very rare events, where a certain fluctuating entropy takes a negative value. Therefore, they are appropriately described with the use of large deviation theory. In particular, here we focus on the GCEM fluctuation theorem for Markov pure jump processes. This relation is written as a symmetry in the large deviation function (or the corresponding scaled cumulant generating function) associated with the probability distribution of the fluctuating entropy and is also known as the GCEM symmetry. 

More broadly, one can consider a class of functionals of a stochastic path known as time-integrated currents. Whereas the entropy, which is a specific time-integrated current, has a symmetric large deviation function, in general other currents do not display this symmetry. The GCEM symmetry has its roots on the fact that  the entropy is a very special functional given by the logarithm of the weight of the path divided by the weight of the time-reversed path. Hence, one question that arises is that if it is possible to find other currents with a symmetric large deviation function and what would be the physical origin of the symmetry.

This  problem has been recently addressed in \cite{barato12} where it was shown that for a restricted class of Markov pure  jump processes a current different from entropy presents a symmetric large deviation function. As  examples, it was shown in \cite{barato12} that besides the entropy, the height in a surface growth model (see also \cite{barato10}) and the mechanical work in a toy model for a molecular motor display such symmetry. More clearly, Markov jump processes are used to describe a large amount of physical processes out of equilibrium and time-integrated currents are important physical observables. Therefore, a more complete theory about symmetries of large deviation functions associated with time-integrated currents might be relevant for the theoretical understanding of nonequilibrium statistical physics.          

We pursue this direction in the present paper where we obtain a necessary condition in order to have the symmetry in a non-entropic current. This condition is phrased in terms of Schnakenberg network theory \cite{schnakenberg76}, where the states of the Markov process form a network, with the transition rates representing the edges and the states the vertices. We show that in order to have a symmetric current different from entropy, the current increments related to cycles in this network have to be degenerate. This result comes from an analysis of the characteristic polynomial of a modified Markov generator \cite{andrieux07,andrieux07',lebowitz99}.   

In addition, we consider explicitly a four-state system with three cycles. We show that in this case two other symmetries, different from the GCEM symmetry, can be found (see Fig. \ref{fig5}). We also demonstrate that in a fully connected four-state system many other symmetries arise. As for the physical origin of the symmetries, following previous work \cite{barato12}, we show that they are also related to time-reversal. However, they come from the time-reversal of a group of trajectories, where the grouping depends on the specific current under consideration and is related to the degeneracies in the increments of the cycles.

We note that links between Schnakenberg network theory and fluctuation relations have been addressed previously in the literature \cite{Jia1,andrieux07,fagionato11,puglisi10}. The main differences between these works and the present paper are: authors in \cite{andrieux07,fagionato11} study multidimensional joint distribution of currents while here we consider the probability distribution of a single non-entropic current; we focus our study on elementary currents (see section 3)  while the work explained in \cite{Jia1}  concerns fluctuating currents associated with topological cycles  (Kalpazidou  \cite{Kal} provided a survey of the interconnection between  topological cycles and edges). Finally, the  effect of the coarse graining procedure on the entropy production and its relation to Schnakenberg network theory is analysed in \cite{puglisi10}.    

The paper is organized in the following way. In the next section we define time-integrated currents and show how their scaled cumulant generating function can be obtained from a modified generator. In section 3 we introduce Schnakenberg network theory, which is used in section 4 where we obtain the necessary condition. Section 5 contains an analysis of the four-state system with three cycles. The new symmetry as a result of the time-reversal of a group of paths is discussed in section 6. We conclude in section 7. Moreover, in the \ref{AB} an extension of the  Kirchhoff's law in the large deviation regime, which is important for Schnakenberg network theory, is presented and the four-state fully connected system is analyzed in \ref{AA}.

\section{Current probability distribution and modified generator}

Pure jump continuous time Markov processes \cite{Fel1} are defined by the transition rates between a pair of states. If we consider a pair of states $(x,x')$, the probability per unit time of going from state $x$ to $x'$ (the transition rate) is denoted by $w_{x\rightarrow x'}$. In this paper we restrict to a finite set of states represented by $\Gamma$. The probability of being in a state $x\in \Gamma$ at time $t$ follows the master equation, which reads
\begin{equation}
\frac{d}{dt}P(x,t)=-\lambda(x) P(x,t)+\sum_{x\neq x'}P(x',t)w_{x'\to x}, 
\end{equation} 
where $\lambda(x)=\sum_{x'\neq x}w_{x\to x'}$ is the escape rate from state $x$. This equation is also normally written in the matrix form 
$\frac{d}{dt}P_t= P_t\mathcal{L} $, where $\mathcal{L}$ is the Markov generator. It is defined as
\begin{equation}	
\mathcal{L}_{x x'}=\left\{
\begin{array}{ll} 
 w_{x\to x'} & \quad \textrm{ if } x'\neq x\\
 -\lambda(x) & \quad \textrm{ if }  x'=x 
\end{array}\right.\,.
\label{generator}
\end{equation}

The object we study here is the so-called time-integrated current (or just current). This is a functional of the stochastic trajectory, which is a sequence of jumps $x(0),x(t_1),\ldots x(t_{M-1}),x(t_M)$, taking place at random times $t_i$, within a fixed time interval $[0,T]$ , where $M$ is the fluctuating total number of jumps in the trajectory. More precisely, the stochastic trajectory starts at state $x(0)$ at time $t=0$ and   then jumps iteratively from state $x(t_{i-1})$ to state $x(t_i)$ at time $t=t_i$ until it reaches a state $x(t_M)$, where it stays until at least $t=T$. Representing a stochastic trajectory by $X_{\left[0,T\right]}$, a time-integrated current is a functional written as 
\begin{equation}
J_{T}[X_{\left[0,T\right]}]=\sum_{i=0}^{M-1}f(x(t_{i}),x(t_{i+1})),
\label{cur}
\end{equation}
where $f(x,x')$ is the increment of the current when the trajectory makes a jump $x\to x'$. Furthermore, a current is a functional such that this increment has the property of being antisymmetric, i.e., $f(x,x')=-f(x',x)$.  

Since the current is a functional of the stochastic trajectory we can consider a probability distribution of currents in the following way. Given a time interval $T$ the probability that the current (\ref{cur}) takes the value $aT$ is written as
\begin{equation}
P\left(j_T=a\right)= \sum_{X_{\left[0,T\right]}}\mathbb{P}(X_{\left[0,T\right]}) \delta(J_T[X_{\left[0,T\right]}]-aT),
\label{probcurrent}
\end{equation}
where $j_T=J_T/T$ is the time-averaged current, $\mathbb{P}(X_{\left[0,T\right]})$ is the weight of the path $X_{\left[0,T\right]}$ and the sum represents an integral over all possible trajectories. In most of this paper we do not carry the explicit dependence of the current on the stochastic trajectory $X_{\left[0,T\right]}$, writing only $J_T$. 

The fluctuation theorem is a symmetry in the probability distribution of the entropy current $S_T$, which is defined by the increment $f(x,x')= \ln \frac{w_{x\to x'}}{w_{x'\to x}}$, i.e.,
\begin{equation}
S_{T}[X_{\left[0,T\right]}]=\sum_{i=0}^{M-1}\ln \frac{w_{x(t_{i})\to x(t_{i+1})}}{w_{x(t_{i+1})\to x(t_{i})}}.
\label{curentropy}
\end{equation}
We point out that we are considering processes such that if $w_{x\to x'}\neq0$ then $w_{x'\to x}\neq0$, otherwise entropy cannot be defined.  
This symmetry, known as the GCEM symmetry, is valid in the limit of $T\to\infty$ and is related to events in which the entropy current considerably deviates from its average. Therefore, it is conveniently written in terms of the large deviation function $I_s(a)$, which is defined by
\begin{equation}
I_s(a)= \lim_{T\to \infty} -\frac{1}{T}\ln P\left(s_T=a\right),
\end{equation}
where $s_T=S_T/T$. Explicitly, the GCEM symmetry reads 
\begin{equation}
I_s(a)-I_s(-a)=-a.
\label{GCEMlarge}
\end{equation}
Note that we are using the subscript $s$ to denote the large deviation function associated with entropy. For a general current of the form (\ref{cur}), we denote the large deviation function by $I(a)$.

Instead of the probability distribution of a current we can work with the associated generating function. Particularly, the scaled-cumulant generating function related to $J_{T}$ is  defined by  
\begin{equation}
\hat{I}(z)\equiv\lim_{T\rightarrow\infty}\frac{1}{T}\ln \sum_{X_{\left[0,T\right]}}\mathbb{P}(X_{\left[0,T\right]})\exp\left(zJ_T[X_{\left[0,T\right]}]\right),
\end{equation}
It follows from the Varadhan theorem \cite{Varadhan,ellis85,hollander,touchette09} that $\hat{I}(z)$  is the Legendre-Fenchel transform of $I(a)$, that is,
\begin{equation}
\hat{I}(z) = \textrm{sup}_{a\in\mathbb{R}}\{za-I(a)\}.
\end{equation} 
Therefore, the GCEM symmetry can also be written as
\begin{equation}
\hat{I}_s(z)= \hat{I}_s(-1-z).
\label{GCEMscaled}
\end{equation}
More generally, for any current proportional to $\frac{S_{T}}{E}$ in the large deviation regime we obtain the symmetry $\hat{I}(z)= \hat{I}(-E-z)$. Moreover, it can be shown that $\hat{I}(z)$ is given by the maximum eigenvalue of a modified generator associated with the current $J_T$ \cite{lebowitz99}. This modified generator is defined as
\begin{equation}	
\mathcal{L}(z)_{x x'}=\left\{\begin{array}{ll} 
 w_{x\to x'}\exp(z f(x,x')) & \quad \textrm{if } x\neq x'\\
 -\lambda(x) & \quad \textrm{if } x=x'
\end{array}\right.\,.
\label{modgenerator}
\end{equation}
Note that this is not a stochastic matrix, but it is still a Perron-Frobenius matrix: it has a unique real maximum eigenvalue which gives $\hat{I}(z)$.    

In this paper we are interested in finding currents following the symmetry $\hat{I}(z)=\hat{I}(-E-z)$ that are different from entropy in the limit of $T\to \infty$ (i.e. not proportional to $s_T$). A sufficient condition for that is a fully symmetric spectrum of eigenvalues \cite{andrieux07,andrieux07'}. In this case, the characteristic polynomial associated with $\mathcal{L}(z)$,  
\begin{equation}
P(z,y)=\det\left(\mathcal{L}(z)-yId\right),
\label{eq:det}
\end{equation}
where $Id$ is the identity matrix, follows the symmetry $P(z,y)= P(-E-z,y)$. Therefore, our objective is to find currents different from entropy in the large deviation regime such that the characteristic polynomial of their associated modified generators is symmetric. As a general result in this direction, in section \ref{sec4} we obtain a necessary condition for a symmetric characteristic polynomial related to a non-entropic current. Before going into that, in the next section we define elementary currents and introduce Schnakenberg network theory, which are important in analyzing the determinant (\ref{eq:det}).

\section{Elementary currents and Network theory}

We now consider the space of states $\Gamma$ as a graph where the vertices are the states and the edges represent the transition rates. Therefore, if the transition rate between two states is zero there is no edge connecting these states. We denote this graph by $G(\Gamma)$.

Given a pair of states $(x,x')$, the elementary fluctuating current from $x$ to $x'$ is written as $J_T(x,x')$. Moreover, an elementary current is such that $f(x,x')=1$ (which implies $f(x',x)=-1$ ) and the increment is zero for all other pairs of states in $G(\Gamma)$. Therefore, the general current (\ref{cur}) can be written as
\begin{equation}
J_T= \frac{1}{2} \sum_{x,x'}f(x,x')J_T(x,x').
\label{curasele}
\end{equation}

An important restriction on the number of independent elementary currents is the finite time Kirchhoff's law \cite{Kirchkoff}, which reads 
\begin{equation}
\sum_{x'}J_{T}(x,x')=\pm1, 0\qquad\textrm{for all $x\in \Gamma$}.
\end{equation} 
This relation comes from the fact that, during a stochastic trajectory, when the system reaches the state $x$, if there is a subsequent jump, the system will leave $x$. Defining $j_T(x,x')= J_T(x,x')/T$, the above relation can be written as
\begin{equation}
\sum_{x'}j_{T}(x,x')=O(\frac{1}{T})\qquad\textrm{for all $x\in \Gamma$}.
\label{kirchhofff}
\end{equation}
This formula is valid for any trajectory (remember that $j_{T}(x,x')$ is a functional of the trajectory  $X_{\left[0,T\right]}$) and in the typical regime  ($T\to\infty$), it gives the usual Kirchhoff's law. As we explain in \ref{AB} it is also valid in the (non typical) large deviation regime, which is normally not appreciated in the literature (see \cite{Maes1} for a counter example).

\begin{figure}
\centering\includegraphics[width=100mm]{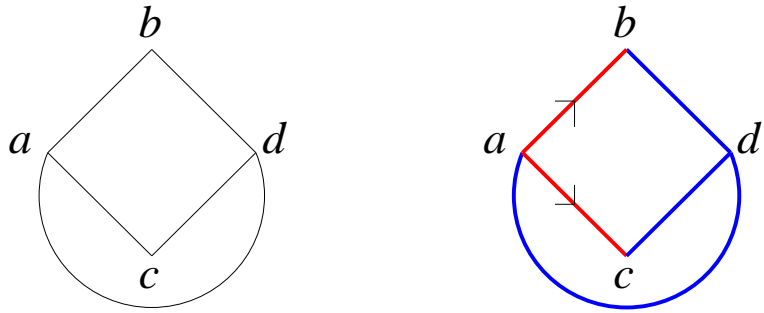}
\caption{On the left, the network of transitions for the four-state and three cycles system, where a link indicates that the transition rates between the pair of states is non-zero. On the right, the spanning tree is in blue and the two chords are in red. The fundamental cycle associated with the chord $(a,b)$ is $\mathcal{C}_{1}=(a,b,d,a)$, while the one related to the chord $(a,c)$ is $\mathcal{C}_2= (a,c,d,a)$.}
\label{fig1}
\end{figure}

Let us now introduce Schnakenberg network theory \cite{schnakenberg76} (see also \cite{andrieux07,puglisi10}). First we introduce the concepts of cycle, fundamental cycle, spanning tree, and chord. A cycle in the network $G(\Gamma)$ is a closed path (or loop): this is a a sequence of jumps $\mathcal{C}=\left[x_{1},x_{2},...,x_{1}\right]$, which finishes in the same state it started and does not go through the same state more than one time. Note that, except for cyclic reordering, the order of the states is relevant.

Given a current of the form (\ref{cur}), the increment related to a cycle is   
\begin{equation}
K(\mathcal{C})\equiv\sum_{i=1}^{n(\mathcal{C})}f(x_{i},x_{i+1}),
\label{Jc}
\end{equation}
where $n(\mathcal{C})$ is the number of states in the cycle. Furthermore, the product of rates of the cycle $\mathcal{C}$, which we refer to as the rate of the cycle, is given by 
\begin{equation}
W(\mathcal{C})\equiv\prod_{i=1}^{n(\mathcal{C})}w_{x_{i}\rightarrow x_{i+1}}.
\end{equation}
The set of all cycles  in $G(\Gamma)$ with at least three jumps is denoted by $\Theta=\left\{ \mathcal{C}/n(\mathcal{C})\geq3\right\}$ (note that $K(\mathcal{C})=0$ if $\mathcal{C}$ is a transposition, a cycle with two states). Given a cycle $\mathcal{C}=\left[x_{1},x_{2},...x_{n(\mathcal{C})},x_{1}\right]$, its reverse $\left[x_{1},x_{n(\mathcal{C})},...x_{2},x_{1}\right]$ is represented by $\overline{\mathcal{C}}$ and, therefore, $K(\overline{\mathcal{C}})=-K(\mathcal{C})$.

\begin{figure}
\centering\includegraphics[width=100mm]{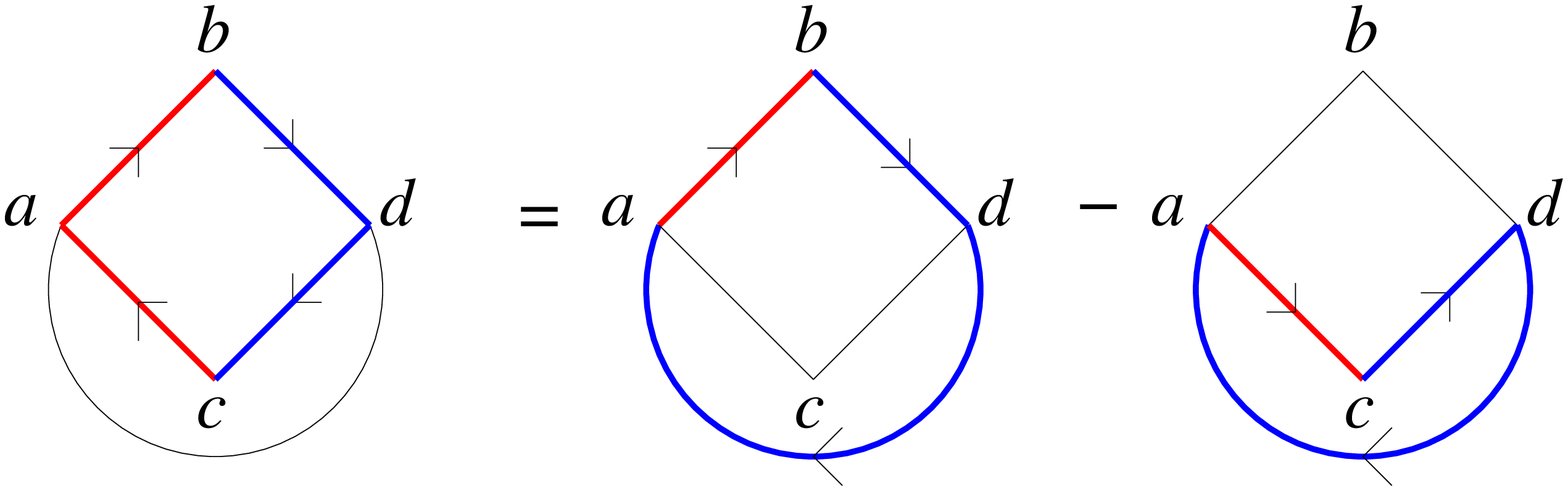}
\caption{Graphical representation of relations (\ref{Krel}) and (\ref{Wrel}). On the left side we have the cycle $\mathcal{C}_3= (a,b,d,c,a)$, the first cycle on the right side is $\mathcal{C}_1=(a,b,d,a)$, and the second is $\mathcal{C}_2=(a,c,d,a)$.}
\label{fig2}
\end{figure}

A spanning tree is a undirected connected set of edges that goes through all the vertices in $G(\Gamma)$ and has no cycles (see Fig. \ref{fig1}). After choosing a maximal spanning tree, all edges that are not part of it are called chords. A chord is represented by $l$ and connects the states $x_l$ and $x_l'$ (from $x_l$ to $x_l'$). The number of chords does not depend on the choice of the maximal tree and is called chord number (or cyclomatic number). Whenever we add a chord to a maximal spanning tree we get a cycle, which is called fundamental cycle $\mathcal{C}_l$. These chords are important because for $T\to\infty$ a general current of the form (\ref{cur}) can be written as a linear composition of the elementary currents through the chords (see formula (\ref{curT}) below). As an example, in Fig. \ref{fig1} we consider a four-state system, labeled by $a,b,c,d$, where there are no transitions between $b$ and $c$. We show a spanning tree where the two links left out are the chords $(a,b)$ and $(a,c)$. Moreover, the two fundamental cycles associated with each chord are displayed in Fig. \ref{fig2}. Note that this network has eight different spanning trees.

Moreover, any cycle in $\Theta$ can be written as a linear combination of the fundamental cycles, i.e., the fundamental cycles form an orthogonal basis (this was originally found by Kirchhoff \cite{Kirchkoff}). In order to see that we define the scalar product between a cycle and a pair of states as:
\begin{equation} 
\fl \left\langle \mathcal{C},(x,x')\right\rangle
=\left\{\begin{array}{ll} 
 0 & \textrm{if $(x,x')$ is not part of $\mathcal{C}$}\\
1 & \textrm{if $(x,x')$ is part of $\mathcal{C}$ and in the same direction}\\	
-1 & \textrm{if $(x,x') $ is part of $\mathcal{C}$ and in the opposite direction}
\end{array}\right.\,.
\label{scalar}
\end{equation}
This quantity gives an answer to the following question: is $(x,x')$ part of the cycle $\mathcal{C}$, and if yes, are they oriented in the same direction? It is possible to show (see \cite{schnakenberg76}) that any cycle $\mathcal{C}$ can be written as 
 \begin{equation}
\mathcal{C}= \sum_l \left\langle
 \mathcal{C},(x_l,x_l') \right\rangle
 \mathcal{C}_l.
\label{cyclesrel}
\end{equation}
The basic reasons for this formula are that a chord is not shared by two fundamental cycles and it must belong to a cycle $\mathcal{C}\in \Theta$. This means that the increment of a cycle  is given by
\begin{equation}
K(\mathcal{C})= \sum_l \left\langle
 \mathcal{C},(x_l,x_l') \right\rangle
 K(\mathcal{C}_l).
\label{Krel}
\end{equation}
Furthermore,  due to the fact that $\ln\left(\frac{W(\mathcal{C})}{W(\overline{\mathcal{C}})}\right)$ is a particular case of a current $K(\mathcal{C})$, as a corollary we get 
\begin{equation}
\frac{W(\mathcal{C})}{W(\overline{\mathcal{C}})}= \prod_l\left(\frac{W(\mathcal{C}_l)}{W(\overline{\mathcal{C}_l})}\right)^{\left\langle \mathcal{C},(x_l,x_l') \right\rangle}.
\label{Wrel}
\end{equation} 

The elementary current through an edge $(x,x')$ can be written as a linear combination of elementary currents through the chords $l$. Explicitly, from the finite time Kirchhoff's law (\ref{kirchhofff}) we obtain the asymptotic relation
\begin{equation}
j_{T}(x,x')=\sum_{l}j_{T}(x_l,x_l')\left\langle \mathcal{C}_{l}, (x,x')\right\rangle+O(\frac{1}{T}).
\label{kirchhoff2}
\end{equation}
Therefore, from (\ref{Jc}) we see that the general current (\ref{curasele})  takes the asymptotic form   
\begin{eqnarray}
j_{T} 
& = & \frac{1}{2}\sum_{x,x'}\sum_{l} f(x,x') j_{T}(x_l,x_l')\left\langle \mathcal{C}_l,(x,x')\right\rangle + O(\frac{1}{T}) \nonumber  \\ 
& = & \frac{1}{2}\sum_{l}  j_{T}(x_l,x_l')\sum_{x,x'} f(x,x') \left\langle \mathcal{C}_{l},(x,x')\right\rangle + O(\frac{1}{T}) \nonumber \\
 & = & \sum_{l}j_{T}(x_l,x_l')K(\mathcal{C}_{l})+O(\frac{1}{T}).
 \label{curT}
\end{eqnarray}
In particular, for the case of  the entropy we have
\begin{equation}
s_T= \sum_l j_{T}(x_l,x_l')\ln\frac{W(\mathcal{C}_l)}{W(\overline{\mathcal{C}}_l)}+O(\frac{1}{T}).
\label{entT}
\end{equation}

Relation (\ref{kirchhoff2}) is normally written only for the (average) stationary current (see \cite{andrieux07,Alt}), however, we stress that it is valid for any stochastic trajectory. An elegant and heuristic proof of (\ref{kirchhoff2}) for the stationary current can be found in \cite{Polettini}, while a rigorous proof is given in \cite{Biggs}.  

A graphical representation of the decomposition of a cycle $\mathcal{C}$ into fundamental cycles using the scalar product (\ref{scalar}) for a four-state system is showed in Fig. \ref{fig2}. The concepts developed here are central in studying the characteristic polynomial (\ref{eq:det}). We proceed by deriving a general necessary condition for the symmetry to set in for a non-entropic current. As we show below, this condition is related to degeneracy in the set of values of the increments of cycles $K(\mathcal{C})$.    

\section{Necessary condition for the symmetry}
\label{sec4}

Using the Leibniz formula for determinants, the characteristic polynomial (\ref{eq:det}) can be written as 
\begin{equation}
P(z,y)=\sum_{\pi\in S(\Gamma)}\textrm{sgn}(\pi)\prod_{x}\left(L(z)_{x,\pi_{x}}-y\delta_{x,\pi_{x}}\right)
\label{eq:devdet}
\end{equation}
where $\pi$ is a permutation of the states $x\in \Gamma$ and $S(\Gamma)$ is the permutation group associated with $\Gamma$.

Moreover, each term of the determinant (\ref{eq:devdet}) can be represented by a graph associated with the permutation $\pi$ on the network of states. Due to the bijectivity of $\pi$, the graph is such that there is one transition rate entering and one transition rate leaving each state of the network. Therefore,  as is well known in discrete mathematics \cite{Bona}, each permutation $\pi$ can be decomposed as product of disjoint cycles  (with no common state). For example, in the case of seven states $\left\{ 1,2,3,4,5,6,7\right\} $, the permutation $\pi(i)=2i$ mod$7$, which is represented by $\left\{ 2,4,6,1,3,5,7\right\}$, can be written as $\left(1,2,4,1\right)\circ\left(3,6,5,3\right)\circ\left(7\right)$. We denote by $\mathbb{C}_\pi$ the subset of  the set of all cycles $\Theta$ obtained from the cycles decomposition associated with $\pi$. For example, in the case of $\pi(i)=2i$ mod$7$ we have $\mathbb{C}_\pi=\left\{ \left(1,2,4,1\right),\left(3,6,5,3\right)\right\}$.  Note that, as a consequence of the disjoint character of the cycles, the order of the cycles in $\mathbb{C}_\pi$ is irrelevant.

The increment related to $\mathbb{C}_\pi$  is written as  
\begin{equation}
K(\mathbb{C}_\pi)=\sum_{\mathcal{C}\in \mathbb{C}_\pi}K(\mathcal{C})
\end{equation}
and the transition rate is given by
\begin{equation}
W(\mathbb{C}_\pi)=\prod_{\mathcal{C}\in \mathbb{C}_\pi}W(\mathcal{C}).
\end{equation}
With these quantities, the characteristic polynomial (\ref{eq:devdet}) takes the expression
\begin{equation}
P(z,y)=\sum_{\pi\in S(\Gamma)}\textrm{sgn}(\pi)\exp\left(zK(\mathbb{C}_\pi)\right)W(\mathbb{C}_\pi)g_{\pi}(y)
\label{dete}
\end{equation}
where $g_{\pi}(y)$ is the term which comes from the fixed states of the permutation ($\pi(x)=x$) and from the transpositions ($\pi(x)=x',$$\pi(x')=x$). Fixed states under the permutation contribute with the term $-(y+\lambda(x))$, while transpositions contribute with the term $w_{x\rightarrow x'}w_{x'\rightarrow x}$. It then follows that
\begin{equation}
g_{\pi}(y)=\prod_{x\in \Gamma/\pi(x)=x}\left(-y-\lambda(x)\right)\prod_{\left\{ x,x'\right\} \in(\Gamma,\Gamma)/\pi(x)=x',\pi(x')=x}\left(w_{x\rightarrow x'}w_{x'\rightarrow x}\right).
\end{equation}

We define the set $\Lambda=\left\{ \mathbb{C}_\pi,\pi\in S(\Gamma)\right\}$ of the different disjoint cycle decomposition related to  $S(\Gamma)$. Equation (\ref{dete}) can be rewritten as  
\begin{equation}
P(z,y)=\sum_{\mathbb{C}\in\Lambda}\exp\left(zK(\mathbb{C})\right)W(\mathbb{C})\sum_{\pi\in S(\Gamma)/\mathbb{C}_\pi=\mathbb{C}}\textrm{sgn}(\pi)g_{\pi}(y),
\label{dete2}
\end{equation}
where the last sum is restricted to permutations $\pi$ such that $\mathbb{C}_\pi=\mathbb{C}$. Since $\mathbb{C}_{\pi^{-1}}=\overline{\mathbb{C}}_{\pi}$, where $\overline{\mathbb{C}}_\pi$ is the set of cycles composed of $\mathbb{C}_\pi$ cycles in the reversed direction, we obtain
\begin{eqnarray}
P(z,y) & = & \sum_{\mathbb{C}\in\Lambda}\exp\left(zK(\overline{\mathbb{C}})\right)W(\overline{\mathbb{C}})\sum_{\pi\in S(\Gamma)/\mathbb{C}_\pi=\overline{\mathbb{C}}}\textrm{sgn}(\pi)g_{\pi}(y)\label{carcy}\nonumber\\
 & = & \sum_{\mathbb{C}\in\Lambda}\exp\left(-zK(\mathbb{C})\right)W(\overline{\mathbb{C}})\sum_{\pi\in S(\Gamma)/\mathbb{C}_{\pi^{-1}}=\mathbb{C}}\textrm{sgn}(\pi)g_{\pi}(y)\nonumber \\
 & = & \sum_{\mathbb{C}\in\Lambda}\exp\left(-zK(\mathbb{C})\right)W(\overline{\mathbb{C}})\sum_{\pi\in S(\Gamma)/\mathbb{C}_\pi=\mathbb{C}}\textrm{sgn}(\pi)g_{\pi}(y),
\label{carcy2}
 \end{eqnarray}
where we used the properties $K(\overline{\mathbb{C}})= -K(\mathbb{C})$, $g_{\pi^{-1}}(x)=g_{\pi}(x)$ and $\textrm{sgn}(\pi^{-1})=\textrm{sgn}(\pi)$.

Imposing the symmetry   
\begin{equation}
P(-E-z,y)=P(z,y)\qquad\textrm{for all $y$},
\end{equation}
and using formulas (\ref{dete2}) and (\ref{carcy2}), we get
\begin{eqnarray}
\sum_{\mathbb{C}\in\Lambda}\exp\left(EK(\mathbb{C})\right)\exp\left(zK(\mathbb{C})\right)W(\overline{\mathbb{C}})\sum_{\pi\in S(\Gamma)/\mathbb{C}_\pi=\mathbb{C}}\textrm{sgn}(\pi)g_{\pi}(y)\nonumber\\=
\sum_{\mathbb{C}\in\Lambda}\exp\left(zK(\mathbb{C})\right)W(\mathbb{C})\sum_{\pi\in S(\Gamma)/\mathbb{C}_\pi=\mathbb{C}}\textrm{sgn}(\pi)g_{\pi}(y).
\label{fondsym}
\end{eqnarray}
We immediately see that, due to  the independence of the family of exponential functions, the different and non-vanishing values of the set of increments  $K(\Lambda)\equiv\{K(\mathbb{C}),\mathbb{C}\in\Lambda \}$ play an important role in the analysis of this equation. In particular, we consider two complementary subsets $\Omega_1=\left\{ \mathbb{C}_\pi,\pi\in S(\Gamma)/\mathbb{C}_\pi=\{\mathcal{C}_l\}\right\}$ for all chords $l$, and $\Omega_2=\Lambda-\Omega_1$. This means that $\Omega_1$ is the subset of group of cycles formed by groups that contain only one of the fundamental cycles. If the sets of increments $K(\Omega_1)$ and $K(\Omega_2)$ fulfill the following conditions:
\begin{itemize}
\item $K(\mathcal{C}_{l})\neq0$ for all fundamental cycles;
\item $K(\mathcal{C}_{l})\neq K(\mathcal{C}_{l'})$ for all pair of chords $l\neq l'$;
\item the set of increments $K(\Omega_1)$ is disjoint with $K(\Omega_2)$.
\end{itemize}
Then relation (\ref{fondsym}) implies that 
\begin{equation}
\exp\left(EK(\mathcal{C}_l)\right)=\frac{W(\mathcal{C}_l)}{W(\overline{\mathcal{C}_l})}\qquad\textrm{for all chords } l.
\label{andrieux}
\end{equation}
In this case, we see from (\ref{curT}) and (\ref{entT}) that $j_T= \frac{1}{E}s_T+ O(\frac{1}{T})$: we have the familiar GCEM symmetry because at large times $j_T$ is just the entropy current (up to a constant). Therefore we arrive at the following conclusion: in order to have a symmetric current asymptotically  different from entropy either at least one of the increments in the set $\{K(\mathcal{C}_{l})\}$ has to be zero or we need degeneracy in the increments of the fundamental cycles, in the sense that at least one of the last two conditions has to be broken. This is a main result of the present paper and we think it should play an important role in developing a more general theory for the symmetries of large deviation function associated with time-integrated currents. 

Two remarks are in time. First we stress that this is a necessary condition on the full spectrum of the modified generator: it could be that there is a situation where a non-entropic current for which this condition is not fulfilled (implying in an asymmetric characteristic polynomial) still has a symmetric maximum eigenvalue (which gives the scaled cumulant generating function). The other remark is a comparison of our work with \cite{andrieux07'}, where the condition (\ref{andrieux}) was obtained as a necessary and sufficient condition for the GCEM symmetry through a similar analysis of the characteristic polynomial (\ref{eq:devdet}). The "single macroscopic current" considered in \cite{andrieux07'} is proportional to the entropy current and here we are interested in non-entropic currents, which might have a symmetric characteristic polynomial precisely when condition (\ref{andrieux}) is not satisfied. Another difference is that in\cite{andrieux07'} a sort of coarse-grained current is considered, where the number of cycles, and not of the elementary jumps, is counted. In this case one takes a derived Markov process which is obtained from the original process by cutting off cycles \cite{Jia1}.     

Even though we found a general necessary condition, this is still far from a sufficient condition. As we show next, in order for a symmetry to set in the transition rates have to fulfill some constraints, which depend on the kind of degeneracy we have among the increments in the set $K(\Lambda)$. In the next section we perform a full analysis of all possible symmetric time-integrated currents different from entropy that arise in a four-state system. Specifically, we consider the network of states of Fig. \ref{fig1}, where all but one pair of states are connected. We show that in this case, two  classes of symmetric currents different from entropy exist, one of them corresponding to the symmetry previously found in \cite{barato12}. Moreover, in the \ref{AA} we perform a similar analysis for the fully connected network, where there are several different classes of symmetric time-integrated currents.  
       
\section{Three cycles and four-state system}
\label{sec5}

The network of Fig. \ref{fig1} has two $3-$jumps cycles and one $4-$jumps cycle. They are
$\mathcal{C}_1= (a,b,d,a)$, $\mathcal{C}_2= (a,c,d,a)$, and $\mathcal{C}_3= (a,b,d,c,a)$, where $\mathcal{C}_3$ is the $4-$jumps cycle. Besides the forward cycles there are also the three backward cycles, which are denoted by $\overline{\mathcal{C}}_1$, $\overline{\mathcal{C}}_2$, and $\overline{\mathcal{C}}_3$. The increment of the cycles is denoted by $K_i$ and the forward (backward) rate by $W_i$ ($\overline{W}_i$), with $i=1,2,3$. Moreover, for later convenience we write the escape rates from states $b$ and $c$ as $\lambda(c)= \lambda_1$ and $\lambda(b)= \lambda_2$.

If we choose the spanning tree showed in Fig. \ref{fig1} the two chords are $(a,b)$ and $(a,c)$, whereas the fundamental cycles are $\mathcal{C}_1$ and $\mathcal{C}_2$, respectively. The rate and increment of the cycle $\mathcal{C}_3$ fulfills the following relations. From (\ref{Krel}) we obtain (see Fig. \ref{fig2})
\begin{equation}
K_3= K_1-K_2,
\end{equation}   
and from (\ref{Wrel}) we have
\begin{equation}
\frac{W_3}{\overline{W}_3}=\frac{W_1\overline{W}_2}{\overline{W}_1W_2}.
\label{Wrelfour}
\end{equation}
Furthermore, from (\ref{curT}), asymptotically, a general current is written as     
\begin{equation}
j_T= K_1 j_T(a,b)+ K_2 j_T(a,c)+O(\frac{1}{T}).
\label{currfour1}
\end{equation} 
For the case of the entropy the above formula becomes
\begin{equation}
s_T=K_s  \left( \ln \frac{W_1}{\overline{W}_1} j_T(a,b)+ \ln \frac{W_2}{\overline{W}_2} j_T(a,c) \right)+O(\frac{1}{T})
\label{entfour1}
\end{equation} 
where $K_s$ is a constant. Hitherto we have defined entropy as the current for which $K_s=1$, in this section we denote $s_T$ any current that is asymptotically proportional  to entropy. Note that in this case the GCEM symmetry (\ref{GCEMscaled}) changes to $\hat{I}_s(z)=\hat{I}_s(-E_s-z)$, where $E_s=1/K_s$ (see \cite{harris07}).

With the above definitions we can now look for the currents with a symmetric characteristic polynomial. As we showed in the previous section, the symmetry implies  the fulfillment of equation (\ref{fondsym}). In the present case this equation takes the form
\begin{eqnarray}
\fl \left(\e^{K_1 (z+E)}\overline{W}_1+\e^{-K_1 (z+E)}W_1\right)(-\lambda_1-y)+\left(\e^{K_{2} (z+E)}\overline{W}_{2}+\e^{-K_2 (z+E)}W_2\right)(-\lambda_2-y)\nonumber \\  
\fl -\left(\e^{K_3 (z+E)}\overline{W}_3+\e^{-K_3 (z+E)}W_3\right)\nonumber\\  
\fl = \left(\e^{K_1z}W_1+\e^{-K_1z}\overline{W_1}\right)(-\lambda_1-y)+\left(\e^{K_2z}W_2+\e^{-K_2z}\overline{W_2}\right)(-\lambda_2-y)\nonumber \\
\fl  -\left(\e^{K_3z}W_3+\e^{-K_3z}\overline{W_3}\right),
\label{symcondfour}
\end{eqnarray}
where we used the fact that the permutations related to $\mathcal{C}_1$ and $\mathcal{C}_2$ have positive sign and the permutation related to $\mathcal{C}_3$ has negative sign. 
We also showed in the previous section that in order to have a symmetric current different from entropy either one of the increments of the fundamental cycles has to be zero, or one of the fundamental cycles must have the same increment as another cycle (fundamental or non-fundamental). In the present case this leaves us with six possibilities:
\begin{itemize}
\item  $K_1=K_2$; 
\item  $K_1=-K_2$; 
\item $K_2=0$ or $K_1=0$;  
\item $K_1=2K_2$, which gives $K_3=K_2$, or $K_2=2K_1$, which gives $K_3=-K_1$.
\end{itemize}

We first consider the family of currents $j_T^{(\alpha)}$, which have the cycle increments   $K_1=K_2  \equiv  K_\alpha$, implying $K_3=0$.  This means that all currents is this family have the following common asymptotic behavior, 
\begin{equation}
j_T^{(\alpha)}= K_\alpha(j_T(a,b)+j_T(a,c))+O(\frac{1}{T}).
\label{cura}
\end{equation}        
Note that the sufficient relation for the symmetry (\ref{symcondfour}) depends only on the values of $K_1$ , $K_2$ and $K_3$. Therefore, it is identical for all currents in this family.
 
We want to find out under which conditions a current with the assymptotic form (\ref{cura}) is symmetric and different from $s_T$. From equation (\ref{symcondfour}) we obtain 
\begin{equation}
\exp(K_\alpha E_\alpha)=\frac{W_1+W_2}{\overline{W}_1+\overline{W}_2}.
\label{E1first}
\end{equation} 
where we are using $E_\alpha$ for the symmetric factor related to current (\ref{cura}). This relation defines the value of $E_\alpha$ as a function of the transition rates.  Moreover, also from (\ref{symcondfour}) we get
\begin{equation}
\frac{W_1+W_2}{\overline{W}_1+\overline{W}_2}=\frac{W_1\lambda_1+W_2\lambda_2}{\overline{W}_1\lambda_1+\overline{W}_2\lambda_2}.
\label{E1second}
\end{equation} 
The transition rates fulfill this equation if $\lambda_1=\lambda_2$ or $\frac{W_1}{\overline{W_1}}=\frac{W_2}{\overline{W_2}}$. Clearly, from equations (\ref{entfour1}) and (\ref{cura}), in the second case we have $j_T^{(\alpha)}= s_T/E_\alpha+ O(\frac{1}{T})$ (with $K_s=1$). Therefore, we obtain that the spectrum related to $j_T^{(\alpha)}$ is symmetric and this current is different from entropy if
\begin{equation}
\lambda_1=\lambda_2\qquad\textrm{and}\qquad\frac{W_1}{\overline{W_1}}\neq\frac{W_2}{\overline{W_2}}.
\label{conda}
\end{equation}
Therefore, we have the symmetry $\hat{I}_\alpha(z)=\hat{I}_\alpha(-E_\alpha-z)$ with $E_\alpha$ given by (\ref{E1first}).

The second family of currents $j_T^{(\beta)}$ corresponds to  the case  $K_1=-K_2  \equiv  K_\beta$, where $K_3=2K_\beta$.  The asymptotic behavior of the currents in this family is 
\begin{equation}
j_T^{(\beta)}= K_\beta(j_T(a,b)-j_T(a,c)) +O(\frac{1}{T}).
\label{curb}
\end{equation} 
Following the same procedure of the previous case, from equation (\ref{symcondfour}) we obtain that the symmetric factor is defined as
\begin{equation}
\exp(K_\beta E_\beta)=\frac{W_1+\overline{W}_2}{\overline{W}_1+W_2},
\label{Ebeta}
\end{equation}
and the transition rates have to fulfill the constraint
\begin{equation}
\frac{W_1+\overline{W}_2}{\overline{W}_1+W_2}=\frac{W_1\lambda_1+\overline{W}_2\lambda_2}{\overline{W}_1\lambda_1+W_2\lambda_2}.
\end{equation}
In this case the transition rates are such that $\lambda_1=\lambda_2$ or $\frac{W_1}{\overline{W_1}}=\frac{\overline{W_2}}{W_2}$, where the second equality implies $j_T^{(\beta)}$ proportional to entropy. In addition, since $K_3$ is different from zero, relation (\ref{symcondfour}) gives us an extra constraint, which is
\begin{equation}
\left(\frac{W_1+\overline{W}_2}{\overline{W}_1+W_2}\right)^2= \frac{W_1\overline{W}_2}{\overline{W}_1W_2},
\end{equation}
where we used $\exp(2K_\beta E)= \frac{W_3}{\overline{W}_3}$ and relation (\ref{Wrelfour}). This equation leads to $W_1\overline{W}_1= W_2\overline{W}_2$. Therefore, we conclude that $\hat{I}_\beta(z)$ is symmetric, with the symmetric factor $E_\beta$ given by (\ref{Ebeta}), and different from entropy if
\begin{equation}
\left\{\begin{array}{l} 
\lambda_1=\lambda_2\\
W_1\overline{W}_1= W_2\overline{W}_2
\end{array}\right.\,
\qquad\textrm{and}\qquad\frac{W_1}{\overline{W}_1}\neq\frac{\overline{W}_2}{W_2}.
\label{condb}
\end{equation}

For the case $K_2=0$, we take $K_3=K_1= K$. Hence, the associated family of  currents have the asymptotic form  $Kj_T(a,b)$. Equation (\ref{symcondfour}) gives the constraint
\begin{equation}
\frac{W_1}{\overline{W}_1}=\frac{W_1\lambda_1+W_3}{\overline{W}_1\lambda_1+\overline{W}_3}.
\end{equation}
This is satisfied only if $\frac{W_1}{\overline{W}_1}=\frac{W_3}{\overline{W}_3}$, which gives $\frac{W_2}{\overline{W}_2}=1$. This leads to the entropy current being proportional to the elementary current through $(a,b)$ and, therefore, there is no new symmetry in this case. The case $K_1=0$ is analogous.
 
Finally, for the case  $K_1=2K_2=K_3=K$, equation (\ref{symcondfour}) gives the constraint
\begin{equation}
\frac{W_2}{\overline{W}_2}=\frac{W_2\lambda_2+W_3}{\overline{W}_2\lambda_2+\overline{W}_3}.
\end{equation}
This is satisfied only if $\frac{W_2}{\overline{W}_2}=\frac{W_3}{\overline{W}_3}$, which leads to $\frac{W_1}{\overline{W}_1}=\left(\frac{W_2}{\overline{W}_2}\right)^{2} $. This last relation implies the associated current being proportional to entropy: there is no new symmetry. The case $K_2=2K_{1}$ is analogous.

\subsection{Example: two sites SEP}

\begin{figure}
\centering\includegraphics[width=100mm]{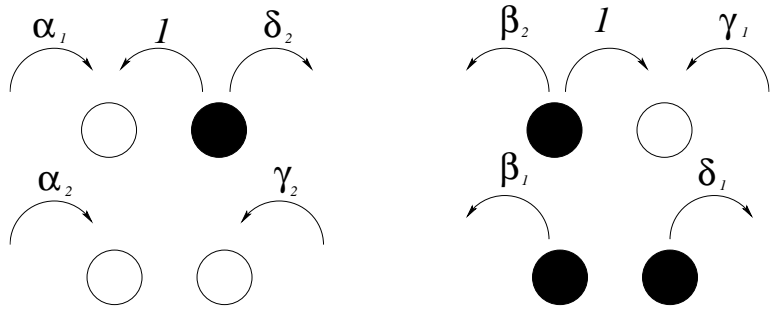}
\caption{Transiton rates for the two sites SEP with the boundary rates depending on the bulk density.}
\label{fig3}
\end{figure}

In order to illustrate these symmetries we consider the symmetric exclusion process (SEP) \cite{derrida93,evans07,derrida07,Chou11}. This is a one dimensional transport model which is driven out of equilibrium by the boundary dynamics. Particles enter and leave the system at the left and right boundary, while they diffuse in the bulk. They also interact in the bulk by imposing that the maximum number of particles per site is one. More generally, we will consider a SEP where the transition rates for particles to enter or leave the system depends on the bulk density. We will restrict our analysis to a two sites system which has the four-state network showed in Fig. \ref{fig1}. The transition rates for this two sites SEP are shown in Fig. \ref{fig3}. Note that, the subscript $1$ ($2$) is related to transitions at the boundary with the other site being occupied (empty). 

Considering Fig. \ref{fig1}, we identify the state where the left (right) site is empty and the right (left) site is occupied with $a$ ($d$) and the state where both sites are occupied (empty) with $b$ ($c$). In the basis $\{a,b,c,d\}$, the generator (\ref{generator}) for the two sites SEP becomes   
\begin{eqnarray}
\mathcal{L}=
\left(
\begin{array}{cccc}
-1-\alpha_1-\delta_2	& \alpha_1  	         & \delta_2             & 1 	\\
\beta_1			&-\beta_1-\delta_1  	 & 0   		        & \delta_1	\\
\gamma_2	        &0	                 & -\gamma_2-\alpha_2   & \alpha_2	\\
1	                &\gamma_1		 & \beta_2		&-1-\beta_2-\gamma_1\\
\end{array}\right).
\label{generatorSEP}
\end{eqnarray}
One physical current that we are interested in is the current between the system and the left reservoir $j_T^{l}$, which is asymptotically equal to minus the current between the system and the right reservoir. This is a functional of the stochastic trajectory such that whenever a particle enters (leaves) the system in the  left boundary the current increases (decreases) by one. Therefore, for the two-sites SEP, the left boundary  current is such that the only non-zero increments are $f(a,b)=1$ and $f(c,d)=1$. Hence, in the long time limit - where $j_T(c,d)=j_T(a,c)+O(\frac{1}{T})$ and $j_T(b,d)=j_T(a,b)+O(\frac{1}{T})$ - we have
\begin{equation}
j_T^{l} \equiv  j_T(a,b)+j_T(c,d)=j_T(a,b)+j_T(a,c)+O(\frac{1}{T}).
\label{left}
\end{equation} 
This current belongs to the family  $j_T^{(\alpha)}$ with $K_\alpha=1$. With the generator (\ref{generatorSEP}) we have $W_1=\alpha_1\delta_1$, $\overline{W}_1=\gamma_1\beta_1$, $W_2=\alpha_2\delta_2$, and $\overline{W}_2=\gamma_2\beta_2$. Therefore, the entropy (\ref{entfour1}), with $K_s=1$, reads 
\begin{equation}
s_T= \ln\frac{\alpha_1\delta_1}{\gamma_1\beta_1}j_T(a,b)+\ln\frac{\alpha_2\delta_2}{\gamma_2\beta_2}j_T(a,c)+O(\frac{1}{T}).
\end{equation}    
An important point is that in general $s_T$ and $j_T^{(l)}$   are not proportional, however for the standard SEP, that is, without dependence of the boundary transitions on the bulk density, the transition rates are such that $j_T^{l}$ and $s_T$ are proportional \cite{bodineau07}.

Furthermore, for the two sites SEP the current from the left reservoir can be divided into two parts: the current when the right site is occupied, which is $j_T(a,b)$, and the current when the right side is empty, which is $j_T(c,d)$. We now consider a second current which is the difference between these two contributions: the current from the left reservoir when the right site is occupied minus the current from the left reservoir when the right site is empty. More clearly, in the large deviation regime this current reads
\begin{equation}
j_T^{ld} \equiv  j_T(a,b)-j_T(c,d)=j_T(a,b)-j_T(a,c)+O(\frac{1}{T}),
\label{leftdiff}
\end{equation}    
which is of the form $j_T^{(\beta)}$ with $K_\beta=1$.

The modified generators for these currents are obtained from (\ref{modgenerator}). As an example, we write down the modified generator for the current (\ref{leftdiff}), which is   
\begin{eqnarray}
\mathcal{L}_\beta(z)=
\left(
\begin{array}{cccc}
-1-\alpha_2-\delta_1	& \alpha_2 \exp(z)  	         & \delta_1             & 1 	\\
\beta_2	\exp(-z)  	&-\beta_2-\delta_2  	 & 0   		        & \delta_2	\\
\gamma_1	        &0	                 & -\gamma_1-\alpha_1   & \alpha_1	\exp(-z)  \\
1	                &\gamma_2		 & \beta_1 \exp(z)		&-1-\beta_1-\gamma_2\\
\end{array}\right).
\label{generator2SEP}
\end{eqnarray}

Our theory predicts that the maximum eigenvalue of the modified generators associated with currents (\ref{left}) and (\ref{leftdiff}) are symmetric, with symmetric factor given by (\ref{E1first}) and (\ref{Ebeta}), respectively,  if some constraints on the transition rates are satisfied. More clearly, if we consider $\alpha_2$ and $\delta_2$ as depending on the other transition rates, from (\ref{conda})  we see that $j_T^l$ has a symmetric and non-entropic scaled cumulant generating function if
\begin{eqnarray}
\alpha_2= \delta_1+\beta_1-\gamma_2 \qquad\textrm{and}\qquad \frac{\alpha_1\delta_1}{\gamma_1\beta_1}\neq\frac{\alpha_2\delta_2}{\gamma_2\beta_2}.
\label{restrictSEP1}
\end{eqnarray}
Moreover, from  (\ref{condb}), for $j_T^{ld}$ the conditions for the symmetry are 
\begin{equation}
\left\{\begin{array}{l} 
\alpha_2= \delta_1+\beta_1-\gamma_2\\
\delta_2= \delta_1 \frac{\alpha_1\beta_1\gamma_1}{\alpha_2\beta_2\gamma_2}
\end{array}\right.\,
\qquad\textrm{and}\qquad\frac{\alpha_1\delta_1}{\gamma_1\beta_1}\neq\frac{\gamma_2\beta_2}{\alpha_2\delta_2}.
\label{restrictSEP2}
\end{equation}
\begin{figure}
\centering\includegraphics[width=100mm]{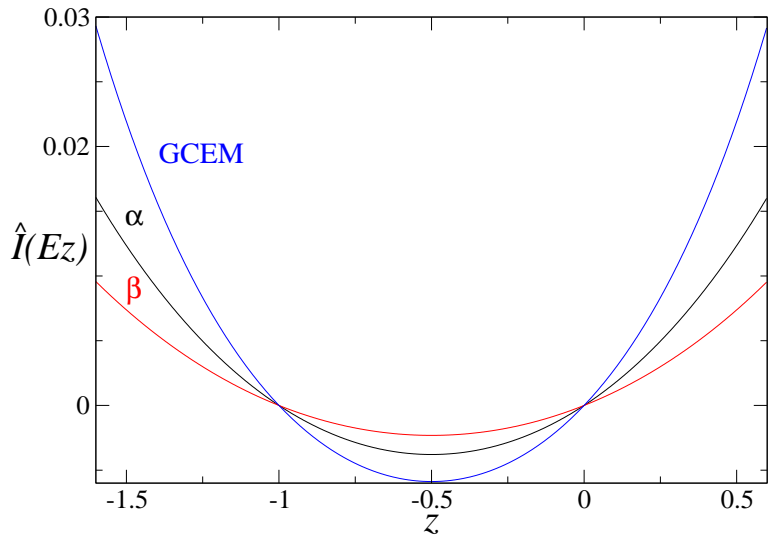}
\caption{Three symmetric currents: $\hat{I}_{\alpha}(z E_\alpha)$ in black, $\hat{I}_{\beta}(z E_\beta)$ in red, and $\hat{I}_s(z)$ in blue. We are considering the currents we defined for the two-site SEP (see text) which corresponds to $K_\alpha=K_\beta=K_s=1$. The value of the transition rates are  
$\alpha_1 = 1$, $\delta_1 = 0.31$,  $\beta_1 = 0.5$, $\gamma_1 = 0.1$, $\gamma_2 = 0.8$, and $\beta_2 = 0.13$. Moreover, $\alpha_2=\delta_1+\beta_1-\gamma_2$ and $\delta_2=\delta_1 (\alpha_1\beta_1\gamma_1)/(\alpha_2\beta_2\gamma_2)$.
}
\label{fig4}
\end{figure}

In Fig. \ref{fig4} we show the scaled cumulant generating function, obtained by numerically calculating the maximum eigenvalue of the modified generator for the case where the constraints (\ref{restrictSEP1}) and (\ref{restrictSEP2}) are satisfied. The scaled cumulant generating functions for the non-entropic currents are appropriately rescaled by their symmetric factors so that they touch the horizontal axis at $-1$.

\begin{figure}
\centering\includegraphics[width=100mm]{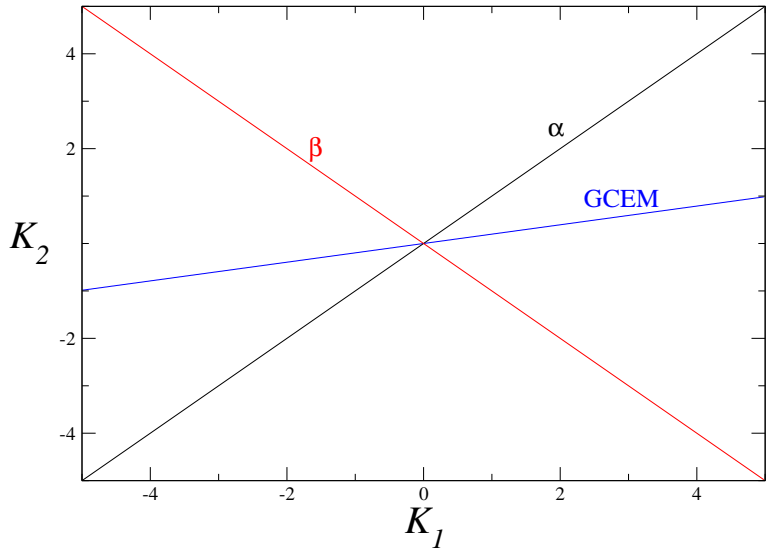}
\caption{The three symmetric currents in the $K_2\times K_1$ plane. The values of the transition rates, which determine the slope of the line for the GCEM symmetry, are $\alpha_1 = 1$, $\delta_1 = 0.31$,  $\beta_1 = 0.5$, $\gamma_1 = 0.1$, $\gamma_2 = 0.8$, and $\beta_2 = 0.13$.
}
\label{fig5}
\end{figure}

Lastly, a very relevant observation is the following. For the present four-state model the space of all possible asymptotic currents in the large deviation regime is two-dimensional: there are two independent elementary currents because the chord number is two. This space of currents can be represented in the $K_{2}\times K_{1}$ plane. In this plane the entropic currents, i.e., the currents that satisfies the GCEM symmetry, are in a line given by 
\begin{equation}
K_{2}= \frac{\ln\frac{W_2}{\overline{W_2}}}{\ln\frac{W_1}{\overline{W_1}}}K_{1}.
\end{equation} 
Our results shows that there are two more lines in this plane with symmetric currents: for the family of currents $j_T^{(\alpha)}$ we have $K_{2}= K_{1}$ and for $j_T^{(\beta)}$ we have $K_{2}= -K_{1}$. Whereas the slope of the line for the GCEM symmetry depends on the transition rates for the other two new symmetries this is not the case. Moreover, different from the GCEM symmetry, the $\alpha$ and $\beta$ symmetries set in only when the transition rates fulfill some constraints. This situation is depicted in Fig. \ref{fig5}. 

\section{Time-reversal as the origin of the symmetries}

As is well known, the GCEM symmetry is a direct consequence of the fact that entropy is the logarithm of the weight of the stochastic path divided by the weight of the reversed path. We argue  that also the new symmetries we found in this paper have their origins in time-reversal. However, for non-entropic currents, this becomes apparent only when we group the trajectories in a certain way. This group of trajectories depends on the current that we are considering and it is determined by the degeneracies of the increments of the cycles. Specifically, we consider the four-state system of the previous section and provide a handwaving  demonstration of the symmetries of the family of currents $j_T^{(\alpha)}$ and $j_T^{(\beta)} $. The presentation here is closely related to reference \cite{barato12}, where the symmetry for the  current $j_T^{(\alpha)}$ was considered. The new results here are the symmetry for the current $j_T^{(\beta)}$, which comes from a different grouping,  and the conjecture that the group of trajectories is determined by the degeneracies in the increment of the cycles. We remark that a (completely different) grouping of trajectories  was also considered in \cite{rahav07}, for the study of the effects of coarse-graining on the fluctuation relations.

Before going into the grouping of paths, let us first present a simple demonstration of the fluctuation theorem. The weight of a stochastic trajectory  $X_{\left[0,T\right]}$ with $M$ jumps is given by  
\begin{equation}
\fl \mathbb{P}(X_{\left[0,T\right]})= \exp[-\lambda(x(t_M))(T-t_M)]\prod_{i=0}^{M-1}w_{x(t_{i})\to x(t_{i+1})}\exp[-\lambda(x(t_i))(t_{i+1}-t_i)].
\label{pathexp}
\end{equation}
The exponential of the waiting times multiplied by the escape rates come from the fact that we are dealing with a continuous time process and we should also multiply this expression by $P(x(t_0),t_0)$ which we are assuming to be uniform for simplicity. The reversed trajectory, where the system starts at state $x(t_M)$ (also with a uniform probability distribution of sates) at time $0$ and jumps from state $x(t_{i+1})$ to state $x(t_{i})$ at time $T-t_{i+1}$, is denoted by $\overline{X_{\left[0,T\right]}} $. Hence,
\begin{equation}
\fl \mathbb{P}(\overline{X_{\left[0,T\right]}})= \exp[-\lambda(x(t_M))(T-t_M)]\prod_{i=0}^{M-1}w_{x(t_{i+1})\to x(t_{i})}\exp[-\lambda(x(t_i))(t_{i+1}-t_i)].
\label{weighttra}
\end{equation}
The entropy current (\ref{curentropy}) is related to the weight of a trajectory divided by the weight of the time-reversed trajectory by
\begin{equation}
\exp(S_T[X_{\left[0,T\right]}])=  \frac{\mathbb{P}(X_{\left[0,T\right]})}{\mathbb{P}(\overline{X_{\left[0,T\right]}})}.
\label{entropypath}
\end{equation}
This is a fundamental relation and it is at the origin of the infinite time fluctuation theorem we consider here as well as all other fluctuation relations. From (\ref{entropypath}) it follows that 
\begin{eqnarray}
\mathbb{P}(X_{\left[0,T\right]}) \exp(-A)\delta(S_T[X_{\left[0,T\right]}]-A)=\mathbb{P}(\overline{X_{\left[0,T\right]}})\delta(S_T[X_{\left[0,T\right]}]-A)\nonumber\\
=\mathbb{P}(\overline{X_{\left[0,T\right]}}) \delta(S_t[\overline{X_{\left[0,T\right]}}]+A)\,,
\end{eqnarray}
where we used the relation $S_T[X_{\left[0,T\right]}]=-S_T[\overline{X_{\left[0,T\right]}}]$. Summing  the above relation over all trajectories and taking relation (\ref{probcurrent}) into account we obtain 
\begin{equation}
\frac{P(S_T=-A)}{P(S_T=A)}=\exp(-A).
\end{equation}
In the limit $T\to \infty$ the above formula implies the GCEM symmetry (\ref{GCEMlarge}). A similar kind of demonstration also holds for currents of the form $j_T^{(\alpha)}$ and $j_T^{(\beta)}$. This happens because a relation analogous to (\ref{entropypath}) is valid for these two families. However, in order to see it we have to consider a functional of an appropriate group of trajectories. The demonstrations that follow are rather heuristic and we plan to provide a precise proof of these symmetries through time-reversal of a group of trajectories in future work.

\subsection{The current $j_T^{(\alpha)}$}

We now consider the current $\tilde{J}_T[X_{\left[0,T\right]}]$, which is defined by the increments
\begin{eqnarray}
f(a,b)=f(a,c)= f(b,d)=f(c,d)= \frac{1}{2} \ln\frac{(W_1+W_2)w_{a\to d}}{(\overline{W}_1+\overline{W}_2)w_{d\to a}}\nonumber\\
f(d,a)= \ln\frac{w_{d\to a}}{w_{a\to d}}.
\label{increa}
\end{eqnarray}
It is easy to check that this current is in the family $j_T^{(\alpha)} $ with $K_\alpha=\ln\frac{(W_1+W_2)}{(\overline{W}_1+\overline{W}_2)}$ (which gives from (\ref{E1first}) $E_\alpha=1$).

Let us now define the group of trajectories. Two trajectories $X_{\left[0,T\right]}$ and $X'_{\left[0,T\right]}$ belong to the same group if they have the same number of jumps taking place at the same times $t_i$. Moreover, if $x(t_i)=a$ then $x'(t_i)=a$, and if $x(t_i)=d$ then $x'(t_i)=d$. On the other hand, if $x(t_i)=b$ or $x(t_i)=c$ then $x'(t_i)=b$ or $x'(t_i)=c$. In this way, a trajectory that stays in states $b$ or $c$ during $m$ of the $M+1$ time intervals is part of a group with $2^{m}$ trajectories. The group of trajectories is denoted by $\{X_{\left[0,T\right]}\}_\alpha$, where the subscript indicates that we are dealing with a current of the type $j_T^{(\alpha)}$. Its weight is just the sum of the weights of all the trajectories in the group, that is,
\begin{equation}
\mathbb{P}(\{X_{\left[0,T\right]}\}_\alpha) =\sum_{X_{\left[0,T\right]} \in \{X_{\left[0,T\right]} \}_{\alpha}}  \mathbb{P}(X_{\left[0,T\right]}).
\label{weightsum}
\end{equation}
Moreover, from the increments (\ref{increa}), we see that the current $\tilde{J}_{T}[X_{\left[0,T\right]}]$ is invariant within the group, meaning that it takes the same value for any $X_{\left[0,T\right]} \in \{X_{\left[0,T\right]} \}_{\alpha}$. Hence, if we want to write this current as a functional of the group of paths $\{X_{\left[0,T\right]} \}_{\alpha}$ we can define
\begin{equation}
\tilde{J}_{T}[\{X_{\left[0,T\right]}\}_\alpha] \, \equiv  \, \tilde{J}_{T}[X_{\left[0,T\right]}]\qquad \textrm{with}\qquad X_{\left[0,T\right]} \in\{X_{\left[0,T\right]}\}_\alpha.
\label{defJt}
\end{equation}
This relation, added to  (\ref{probcurrent}) and (\ref{weightsum}), leads to
\begin{equation}
P\left(\tilde{J}_T=A\right)= \sum_{\{X_{\left[0,T\right]}\}_\alpha}\mathbb{P}(\{X_{\left[0,T\right]}\}_\alpha)  \delta(J_T[\{X_{\left[0,T\right]} \}_{\alpha}]-A).
\end{equation}
Furthermore, the expression for the weight of a group of trajectories $\mathbb{P}(\{X_{\left[0,T\right]}\}_\alpha)$ can be written as
\begin{equation}
\fl \mathbb{P}(\{X_{\left[0,T\right]}\}_\alpha)  = \prod_{i=1}^{M-1} \sum_{x(t_i)\in \{x(t_i)\}_\alpha} w_{x(t_{i})\to x(t_{i+1})}\exp[-\lambda(x(t_i))(t_{i+1}-t_i)],
\end{equation}
where the sum $\sum_{x(t_i)\in \{x(t_i)\}_\alpha}$ has only one term for $x(t_i)=a,d$  and is over the states $b$ and $c$ if the paths in the group have $x(t_i)= b,c$. Similarly we denote by $\mathbb{P}(\{\overline{X_{\left[0,T\right]}}\}_\alpha)$ the  weight of the group formed by the reversed trajectories. Then, the increments of $\ln\frac{\mathbb{P}(\{X_{\left[0,T\right]}\}_\alpha)}{\mathbb{P}(\{\overline{X_{\left[0,T\right]}}\}_\alpha)}$, for the case $x(t_i)=b,c$,  can be written as 
\begin{eqnarray}
\fl\ln\frac{w_{x(t_{i-1})\to b}w_{b\to x(t_{i+1})}\exp[-\lambda(b)(t_{i+1}-t_{i})]+w_{x(t_{i-1})\to c}w_{c\to x(t_{i+1})} \exp[-\lambda(c)(t_{i+1}-t_{i})]}{w_{x(t_{i+1})\to b}w_{b\to x(t_{i-1})}     \exp[-\lambda(b)(t_{i+1}-t_{i})]+w_{x(t_{i-1})\to c}w_{c\to x(t_{i+1}) } \exp[-\lambda(c)(t_{i+1}-t_{i})]}.\nonumber\\
\label{inc}
\end{eqnarray}
In opposition to the ratio of single paths, the time-dependence in the exponential waiting time distribution does not cancel out in general. However, for the case $\lambda(b)=\lambda(c)$, which is condition (\ref{conda}), the increment (\ref{inc}) becomes  a time-independent term which reads  
\begin{eqnarray}
\ln\frac{w_{x(t_{i-1})\to b}w_{b\to x(t_{i+1})}+w_{x(t_{i-1})\to c}w_{c\to x(t_{i+1})}}{w_{x(t_{i+1})\to b}w_{b\to x(t_{i-1})}+w_{x(t_{i-1})\to c}w_{c\to x(t_{i+1})}}
\label{boring}
\end{eqnarray}
Therefore, when condition (\ref{conda}) is fulfilled, with the choice of increments (\ref{increa}),  we expect that asymptotically 
\begin{equation}
\exp(\tilde{J}_{T}[\{X_{\left[0,T\right]}\}_\alpha])=\frac{\mathbb{P}(\{X_{\left[0,T\right]}\}_\alpha)}{\mathbb{P}(\{\overline{X_{\left[0,T\right]}}\}_\alpha)}+O(\frac{1}{T}).
\end{equation}
This means that despite the fact of the current $\tilde{J}$ being different from entropy,  when paths are grouped appropriately a relation analogous to (\ref{entropypath}) in this new coarse-grained space (of trajectories) is valid. Finally, summing over all possible groups of trajectories we obtain 
\begin{equation}
\frac{P(\tilde{J}_T=-A)}{P(\tilde{J}_T=A)}=\exp(-A),
\end{equation}
which proves the symmetry of the current $\tilde{J}$ (and then of all the currents of the family $j_T^{(\alpha)}$). 

This demonstration has two important features. First, the grouping of paths is such that paths in the same group differ by cycles that have the same increment and that is the reason why the current under consideration remains invariant within the group, as in relation (\ref{defJt}). The second feature is that the constraint on the transition rates (\ref{conda}) has the effect of making the ratio (\ref{inc}) time-independent. As we see next these two properties are also shared by the class $j_T^{(\beta)}$.    

\subsection{The current $j_T^{(\beta)}$}

The grouping of trajectories is more complicated in this case and we have to use a different strategy to build an heuristic demonstration. We take the current $\hat{J}_T[X_{\left[0,T\right]}]$, which is defined by the increments
\begin{eqnarray}
f(a,b)=f(a,c)= f(b,d)=f(c,d)= \frac{1}{2}\ln\frac{W_1+\overline{W}_2}{\overline{W}_1+W_2}\nonumber\\
f(d,a)= 1.
\label{increb}
\end{eqnarray}
This is a current of the from (\ref{curb}) with $K_\beta= \ln\frac{W_1+\overline{W}_2}{\overline{W}_1+W_2}$ (implying in $E_\beta=1$). 

In order to roughly define the group $\{X_{\left[0,T\right]} \}_\beta$ let us consider a three jumps sequence between times $[t_{i-1},t_{i+2}]$. For two trajectories in the same group we assume that the first and last state in this three jumps term are the same in both trajectories. However, the two middle states can be different. In order to group cycle $\mathcal{C}_1$ with $\overline{\mathcal{C}}_2$ and $\mathcal{C}_2$ with $\overline{\mathcal{C}}_1$, we choose the following four jumps sequences to get grouped together: 
\begin{eqnarray}
(a,b,d,a)\qquad\textrm{with}\qquad (a,d,c,a)\nonumber\\
(a,d,b,a)\qquad\textrm{with}\qquad (a,c,d,a)\nonumber\\
(d,b,a,d)\qquad\textrm{with}\qquad (d,a,c,d)\nonumber\\
(d,a,b,d)\qquad\textrm{with}\qquad (d,c,a,d).
\label{eq3jump}
\end{eqnarray}
For example, this means that if we look at this three jumps piece of two trajectories in the same group, for one trajectory we could have $(a,b,d,a)$ and for the other $(a,d,c,a)$. The reason for the more complicated grouping in relation to the previous case is that two paths in the same group differs not only by a state $c$ instead of a state $b$, but also the order of the middle states in the three jumps sequence is inverted. 

Similarly to the previous case we expect that the current $\hat{J}_T[X_{\left[0,T\right]}]$ is invariant within the group, i.e.,
\begin{equation}
\hat{J}_{T}[\{X_{\left[0,T\right]}\}_\beta] \, \equiv  \, \hat{J}_{T}[X_{\left[0,T\right]}]\qquad \textrm{with} X_{\left[0,T\right]} \in\{X_{\left[0,T\right]}\}_\beta.
\label{defJtb}
\end{equation}
We define ${\mathbb{P}(\{X_{\left[0,T\right]}\}_\beta)}$ as the sum of the weights in the group. Let us take a three jumps term in the sum $\ln\frac{\mathbb{P}(\{X_{\left[0,T\right]}\}_\beta)}{\mathbb{P}(\{\overline{X_{\left[0,T\right]}}\}_\beta)}$, such that, for example, we have either the cycle $(a,b,d,a)$ or the cycle $(a,d,c,a)$ in the trajectories in the group. This piece of trajectory should contribute with
\begin{equation}
\frac{W_1\exp(-\lambda(b)\Delta t_1-\lambda(d)\Delta t_2)+\overline{W}_2\exp(-\lambda(d)\Delta t_1-\lambda(c)\Delta t_2)}{W_2\exp(-\lambda(c)\Delta t_1-\lambda(d)\Delta t_2)+\overline{W}_1\exp(-\lambda(d)\Delta t_1-\lambda(b)\Delta t_2)},
\end{equation}
where $\Delta t_1= t_{i+1}-t_{i}$ and $\Delta t_2= t_{i+2}-t_{i+1}$. If the first condition in (\ref{condb}), which is $\lambda(b)=\lambda(c)$, is satisfied the above term becomes 
\begin{equation}
\frac{W_1\exp(-\lambda(b)\Delta t_1-\lambda(d)\Delta t_2)+\overline{W}_2\exp(-\lambda(d)\Delta t_1-\lambda(b)\Delta t_2)}{W_2\exp(-\lambda(b)\Delta t_1-\lambda(d)\Delta t_2)+\overline{W}_1\exp(-\lambda(d)\Delta t_1-\lambda(b)\Delta t_2)}.
\label{term}
\end{equation}
In order to get a time-independent term  we need the second condition in (\ref{condb}) to be fulfilled. Explicitly, for $W_1\overline{W}_1=W_2\overline{W}_2$ the term (\ref{term}) becomes $\frac{W_1}{W_2}$. Noting that when $W_1\overline{W}_1=W_2\overline{W}_2$, apart from $f(a,d)$, the increments (\ref{increb}) are equal to $\frac{1}{2}\ln\frac{W_1}{W_2}$  we claim that 
\begin{equation}
\exp(\hat{J}_{T}[\{X_{\left[0,T\right]}\}_\beta])=\frac{\mathbb{P}(\{X_{\left[0,T\right]}\}_\beta)}{\mathbb{P}(\{\overline{X_{\left[0,T\right]}}\}_\beta)}+O(\frac{1}{T}).
\end{equation}
After multiplying by the delta functional and summing over all groups of trajectories we get the final result,
\begin{equation}
\frac{P(\hat{J}_T=-A)}{P(\hat{J}_T=A)}=\exp(-A).
\end{equation}  

Once again, the degeneracy of the increment of the cycles determines the group of trajectories and the constraints on the transition rates have the effect of making the increments in the sum $\ln\frac{\mathbb{P}(\{X_{\left[0,T\right]}\}_\beta)}{\mathbb{P}(\{\overline{X_{\left[0,T\right]}}\}_\beta)}$ time-independent. We conjecture that whenever the full characteristic polynomial (\ref{eq:devdet}) associated with a non-entropic current is symmetric, the symmetry comes from the time-reversal of a group of trajectories and the group has these two properties.

\section{Conclusions}

In this paper we have studied the symmetries of large deviation functions associated with non-entropic currents in pure jump processes. The most general result we obtained is the necessary condition for the appearance of a symmetric current different from entropy. This condition is related to degeneracies in the increments of the fundamental cycles of Schnakenberg network theory and we believe that it is a good starting point for the development of a general theory for the symmetries of large deviation functions associated with currents.

As an example, we studied four-state systems, where symmetric non-entropic currents were found. In this case we saw that these symmetries set in when the increments of cycles are degenerate and when the transition rates fulfill a set of constraints, which depends on the currents under consideration. Moreover, with this example  we learned an important lesson: the symmetries in non-entropic currents come from the time-reversal of a group of trajectories. The degeneracies in the cycles increments determine the group of trajectories, which is such that trajectories in the same group differ only by cycles with the same increment. Furthermore, the constraints on the transition rates have the effect of making the logarithm of the sum of weights of the paths in a group divided by the sum of weights of the the reversed paths time-independent.  

The demonstration of the new symmetries as the time-reversal of a group of trajectories we presented here is still handwaving and restricted to the four-state and three cycles system. We plan to develop a precise proof in future work by considering a coarse grained space of trajectories where the (exponential of the) non-entropic symmetric current becomes the ratio of the weight of the path and the weight of the time-reversed path. Another interesting direction for future work would be to find situations where the characteristic polynomial of the modified generator is not symmetric but the minimum eigenvalue is still symmetric. It could be that in such case, if it exists, the symmetry in the large deviation function is associated with the time-reversal of some most probable trajectory (or set of trajectories), which would dominate in a sum over all trajectories with a given constraint.

{\noindent \textbf{Acknowledgements}}\newline 
We thank Haye Hinrichsen and David Mukamel for helpful discussions. We also thank Boris Lander and David Abreu for carefully reading the manunscript. ACB is thankful to the Laboratoire J. A. Dieudonn\'e for hospitality.

\appendix
\section{Generalization of Kirchhoff's law}
\label{AB}
Keeping the dependence on the stochastic trajectory $X_{\left[0,T\right]}$  explicit,  equation (\ref{kirchhofff}) reads
\begin{equation}
\sum_{x'}j_{T}(x,x')[X_{\left[0,T\right]}]=O(\frac{1}{T}).
\label{kirchhoffapp}
\end{equation}
We would like to remark two points about this relation. 
\begin{itemize}
\item First, and less generally, for the typical behavior, we get the usual Kirchhoff's law for the steady state current. By ergodicity,  the typical behavior of  $j_{T}(x,x')[X_{\left[0,T\right]}]$ is given by
\begin{equation}  
j_{T}(x,x')[X_{\left[0,T\right]}]\rightarrow J_{inv}(x,x')\equiv\rho_{inv}(x)w_{x\rightarrow x'}-\rho_{inv}(x')w_{x'\rightarrow x},
\label{ss}
\end{equation} 
where $\rho_{inv}$ and  $J_{inv}$ are the mean density and current in the invariant steady state. In this case (\ref{kirchhoffapp}) implies  
\begin{equation}   
\sum_{x'}J_{inv}(x,x')=0,
\label{kirchhoff}
\end{equation} 
which is the usual Kirchhoff's law.

\item   More generally, in the (non typical) large deviation regime,  the  finite time Kirchhoff's law (\ref{kirchhoffapp}) imply that the large deviation function of the joint elementary current is infinite if the usual Kirchhoff's law  is not fulfilled \cite{Maes1}. That means, if  we define the probability distribution of the family $\left\{ j_{T}(x,x'), \left(x,x'\right)\in\Gamma\times\Gamma\right\}$ as
\begin{equation}
P(\{j\}=\{a\})= \sum_{X_t}{\mathbb{P}(X_{\left[0,T\right]})}\prod_{x,x'}\delta\left(j_{T}(x,x')-a_{xx'}\right),
\end{equation}
we have that the large deviation function associated with it respects the relation
\begin{equation} 
I(\left\{ a\right\} )=\infty \;\; \mbox{if}\; \exists x \;\; \mbox{such\; that} \;\; \sum_{x'}a_{xx'} \neq 0.
\end{equation}  
In other words, the  distribution of elementary currents around the atypical limit which do not fulfill Kirchhoff's law decreases faster than exponentially  (and is not considered in this article). 
\end{itemize}

\section{Fully connected four-state network}
\label{AA}
\begin{figure}
\centering\includegraphics[width=100mm]{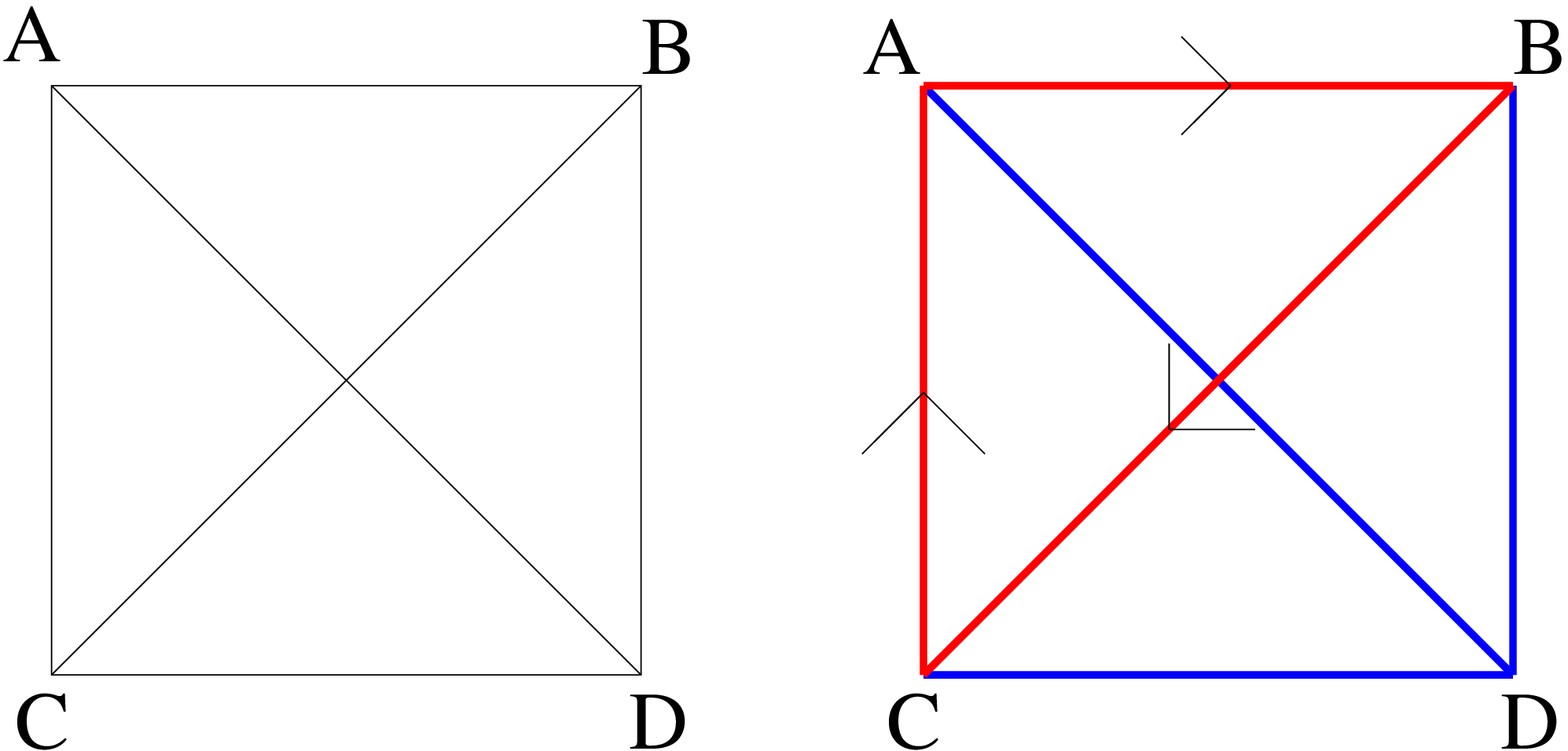}
\caption{Fully connected network of states. On the right, the spanning three is in blue and the chords in red.}
\label{fig6}
\end{figure}

We now turn to the fully connected four-state network, which is shown in  Fig. \ref{fig6}. The states are now denoted be $A$, $B$, $C$ and $D$. The set $\Theta$ contains $14$ cycles, which are:  
\begin{itemize}
\item four cycles with three states: $\mathcal{C}_{1}=(A,B,D,A)$, $\mathcal{C}_{2}=(A,D,C,A)$,$\mathcal{C}_{3}=(B,C,D,B)$, and $\mathcal{C}_{4}=(A,C,B,A)$;
\item three cycles with four states: $\mathcal{C}_{5}=(A,B,C,D,A)$,$\mathcal{C}_{6}=(A,B,D,C,A)$,$\mathcal{C}_{7}=(A,D,B,C,A)$;
\item and the respective reversed cycles.
\end{itemize}
Taking the spanning tree of Fig \ref{fig6} the three chords are $(A,B)$, $(C,A)$ and $(B,C)$. Whereas the three fundamental cycles are $\mathcal{C}_1$, $\mathcal{C}_2$ and $\mathcal{C}_3$, respectively. The increments of the four $3-$states cycles are denoted by $K_i$, with $i=1,2,3,4$. On the other hand, the increments of the three four-state cycles are denoted by $L_i$, with $i=5,6,7$. The forward (backward) rates are represented by $W_i$ ($\overline{W_i}$), for $i=1,2,3,4,5,6,7$. Moreover, we define the escape rates as 
\begin{equation}
\lambda_{1}\equiv\lambda(C),\lambda_{2}\equiv\lambda(B),\lambda_{3}\equiv\lambda(A),\lambda_4\equiv\lambda(D).
\end{equation}

We now have three independent elementary currents and, from equation (\ref{curT}), a general current has the asymptotic form 
\begin{equation}
j_T= K_1j_T(A,B)+K_2j_T(C,A)+K_3j_T(B,C)+O(\frac{1}{T}),
\end{equation}
which implies
\begin{equation}
\fl s_T= K_s\left(\ln\frac{W_1}{\overline{W}_1}j_T(A,B)+\ln\frac{W_2}{\overline{W}_2}j_T(C,A)+\ln\frac{W_2}{\overline{W}_2}j_T(B,C)\right)+O(\frac{1}{T}).
\end{equation}
Furthermore, from relation (\ref{Krel}) we have
\begin{eqnarray}
K_4=-K_1-K_2-K_3\nonumber\\
L_5=K_1+K_3\nonumber\\
L_6=K_1+K_2\nonumber\\
L_7=K_2+K_3, 
\end{eqnarray}
and relation (\ref{Wrel}) gives
\begin{eqnarray}
\frac{W_4}{\overline{W}_4}=\frac{\overline{W}_1\overline{W}_2\overline{W}_3}{W_1W_2W_3}\nonumber\\
\frac{W_5}{\overline{W}_5}=\frac{W_1W_3}{\overline{W}_1\overline{W}_3}\nonumber\\
\frac{W_6}{\overline{W}_6}=\frac{W_1W_2}{\overline{W}_1\overline{W}_2}\nonumber\\
\frac{W_7}{\overline{W}_7}=\frac{W_2W_3}{\overline{W}_2\overline{W}_3}. 
\label{Wrelfully}
\end{eqnarray}
Because in this case  we have more cycles, there is a larger number of possibilities of grouping cycles with the same increment, and different groupings generate different restrictions in the transition rates: each specific grouping has to be treated separately. In the following we show two different cases where a symmetry different from the GCEM symmetry sets in.

First we consider $K_1=K_2 \equiv  K_\gamma$ and $K_3 \, \equiv \, M_\gamma$. This gives a family of currents  $j_T^{(\gamma)}$, which have the asymptotic behavior
\begin{equation}
j_T^{(\gamma)}= K_\gamma j_T(A,B)+K_\gamma j_T(C,A)+M_\gamma j_T(B,C)+O(\frac{1}{T}).
\label{jgamma1}
\end{equation}
Therefore, the set of increments becomes
\begin{equation}
\fl \{K_1,K_2,K_3,K_4,L_5,L_6,L_7\}=\{K_\gamma,K_\gamma,M_\gamma,-2K_\gamma-M_\gamma,K_\gamma+M_\gamma,2K_\gamma,K_\gamma+L_\gamma\},  
\end{equation}
meaning that also cycles $\mathcal{C}_5$ and $\mathcal{C}_7$ have the same increment. Depending on the values of $K_\gamma$ and $M_\gamma$ the grouping of cycles might change. We would like to consider the case where only $K_1=K_2$, $L_5=L_7$ and all other increments are different. In order for this to be true we need the following restrictions on $K_\gamma$ and $L_\gamma$ to be fulfilled,
\begin{eqnarray}
\fl K_\gamma\neq0\qquad M_\gamma\neq0\nonumber\\
\fl |K_\gamma|\neq|M_\gamma|\qquad2|K_\gamma|\neq|M_\gamma|\nonumber\\
\fl 3K_\gamma+M_\gamma\neq 0\qquad4K_\gamma+M_\gamma\neq 0\qquad 3K_\gamma+2M_\gamma\neq0\qquad 2M_\gamma+K_\gamma\neq 0.
\label{condincre}
\end{eqnarray}  
If all these constraints on the increments are satisfied, the equality (\ref{fondsym}) leads to the following relations. Similar to the way we obtained equations (\ref{E1first}) and (\ref{E1second}), if we consider cycles $\mathcal{C}_1$ and $\mathcal{C}_2$ we get
\begin{equation}
\exp(K_\gamma E_\gamma)= \frac{W_1+W_2}{\overline{W}_1+\overline{W}_2}=
\frac{W_1\lambda_1+W_2\lambda_2}{\overline{W}_1\lambda_1+\overline{W}_2\lambda_2}.
\label{Egamma}
\end{equation}
The first equality defines the value of $E_\gamma$, while, as in equation (\ref{E1second}), the second equality is fulfilled for a current different from entropy only if $\lambda_1=\lambda_2$. Considering the cycle $\mathcal{C}_6$ we obtain
\begin{equation}
\exp(2K_\gamma E_\gamma)=\frac{W_1W_2}{\overline{W}_1\overline{W}_2}=\left(\frac{W_1+W_2}{\overline{W}_1+\overline{W}_2}\right)^2,
\end{equation}
where the second equality comes from the first equality in (\ref{Egamma}). This is satisfied if $W_1\overline{W}_2=\overline{W}_1W_2$ or $W_1\overline{W}_1=W_2\overline{W}_2$, where for the first case $J_T^{(\gamma)}$ becomes entropy. Moreover, the cycle $\mathcal{C}_3$ gives
\begin{equation}
\exp(M_\gamma E)=\frac{W_3}{\overline{W}_3},
\end{equation}
which, with relations (\ref{Wrelfully}), (\ref{Egamma}) and the constraint $W_1\overline{W}_1=W_2\overline{W}_2$  gives 
\begin{equation}
\fl j_T^{(\gamma)}=K_\gamma\left(j_T(A,B)+j_T(C,A)+\left(\ln\frac{W_1}{\overline{W}_2}\right)^{-1}\ln\frac{W_3}{\overline{W}_3}j_T(B,C)\right) +O(\frac{1}{T}).
\label{jgamma}
\end{equation}
Finally, the equations obtained from the increments of the cycle $\mathcal{C}_4$ does not bring any new constraints while the cycles $\mathcal{C}_5$ and $\mathcal{C}_7$ give
\begin{equation}
\exp(K_\gamma E_\gamma+M_\gamma E_\gamma)=\frac{W_5+W_7}{\overline{W}_5+\overline{W}_7}.
\end{equation} 
This equation leads to the additional constraint $\frac{W_1}{\overline{W}_2}\frac{W_3}{\overline{W}_3}= \frac{W_5+W_6}{\overline{W}_5+\overline{W}_6}$.
Hence, we conclude that the (non-entropic) family of currents $\left\{j_T^{(\gamma)} \right\}$ has a symmetric large deviation function, with symmetric factor $E_\gamma$ given by (\ref{Egamma}), if the following constraints on the transition rates are respected:
\begin{equation}
\left\{\begin{array}{l} 
\lambda_1=\lambda_2\\
W_1\overline{W}_1= W_2\overline{W}_2\\
\frac{W_1}{\overline{W}_2}\frac{W_3}{\overline{W}_3}= \frac{W_5+W_6}{\overline{W}_5+\overline{W}_6}
\end{array}\right.\,
\qquad\textrm{and}\qquad\frac{W_1}{\overline{W}_1}\neq\frac{\overline{W}_2}{W_2}.
\label{condforgamma}
\end{equation}

Let us make two remarks. The first is that in the above example we grouped cycles $\mathcal{C}_1$ and $\mathcal{C}_2$. However we could have chosen any pair from the four $3-$states cycles, where each pair will give a different family of currents. Therefore, we have six different families: the one we treated and five more. In order to obtain the constraints for the other five one just have to follow the same procedure as above. The second remark is that if we break one of the conditions in (\ref{condincre}) the degeneracies (and also the restrictions in the transition rates) change. For example, if we consider the case $K_\gamma=2L_\gamma$ we have a set of increments given by $\{K_\gamma,K_\gamma,2K_\gamma,-4K_\gamma,3K_\gamma,2K_\gamma,3K_\gamma\}$ and (\ref{jgamma1}) becomes
\begin{equation}
j_T^{(\gamma)}= K_\gamma(j_T(A,B)+j_T(C,A)+j_T(B,C))+O(\frac{1}{T}).
\end{equation}  
Since now cycles $\mathcal{C}_3$ and $\mathcal{C}_5$ are grouped together the restrictions on the transition rates are different from (\ref{condforgamma}). More precisely, it can be shown that for $M_\gamma=2K_\gamma$, besides the restrictions (\ref{condforgamma}), we also need
\begin{equation}
\frac{W_3}{\overline{W}_3}=\frac{W_1W_2}{\overline{W}_1\overline{W}_2},
\end{equation} 
in order to have the symmetry.

As a second case we consider $K_1=-K_2=K_\delta$ and $K_3=M_\delta$. This leads to the set of increments $\{K_\delta,-K_\delta,M_\delta,K_\delta,K_\delta+M_\delta,0,-K_\delta+M_\delta\}$ and the currents of this family have the asymptotic behavior
\begin{equation}
j_T^{(\delta)}= K_\delta j_T(A,B)-K_\delta j_T(C,A)+M_\delta j_T(B,C)
\end{equation}
Similarly to the conditions (\ref{condforgamma}), in order not to get any extra degeneracies the following restrictions on the increments  have to be satisfied,
\begin{eqnarray}
K_\delta\neq0\qquad M_\delta\neq0\nonumber\\
|K_\delta|\neq|M_\delta|\qquad 2|K_\delta|\neq|M_\delta|\qquad|K_\delta|\neq 2|M_\delta|.
\end{eqnarray}
We shall not present all the calculations for this case: they are similar to other calculations in section \ref{sec5} and in this appendix. In the case that all the above conditions are fulfilled we obtain the following. Currents of the form
\begin{equation}
\fl j_T^{(\delta)}=K_\delta\left(j_T(A,B)-j_T(C,A)+\left(\ln\frac{W_1}{W_2}\right)^{-1}\ln\frac{W_3W_2}{\overline{W}_3\overline{W}_1}j_T(B,C)\right)+ O(\frac{1}{T}),
\label{jdelta}
\end{equation}
have a symmetric large deviation function, with symmetric factor given by 
\begin{equation}
\exp(K_\delta E_\delta)= \frac{W_1+\overline{W}_2}{\overline{W}_1+W_2},
\end{equation}
if the following restrictions on the transition rates are fulfilled:
\begin{equation}
\left\{\begin{array}{l} 
\lambda_1=\lambda_2\\
\lambda_3=\lambda_4\\
W_1\overline{W}_1= W_2\overline{W}_2\\
\frac{W_1}{\overline{W}_2}\frac{W_3}{\overline{W}_3}= \frac{W_3+\sqrt{W_5W_6W_7}}{\overline{W}_3+\sqrt{\overline{W}_5\overline{W}_6\overline{W}_7}}
\end{array}\right.\,
\qquad\textrm{and}\qquad\frac{W_1}{\overline{W}_1}\neq\frac{\overline{W}_2}{W_2}.
\label{condfordelta}
\end{equation}

This fully connected case gives us the following intuition. As soon as the network of states has a more sophisticated topology, with a larger number of chords, many other symmetric currents can be found. Therefore, we think that the finding of all possible symmetric currents, by using the methods of Sec. \ref{sec5}, becomes extremely difficult for larger and more complicated networks.

\end{document}